\documentclass{ws-mpla}

\usepackage{amsmath}

\begin{document}

\newcommand\beq{\begin{equation}}
\newcommand\eeq{\end{equation}}
\newcommand\bea{\begin{eqnarray}}
\newcommand\eea{\end{eqnarray}}
\newcommand\bseq{\begin{subequations}} %solo con amsmath
\newcommand\eseq{\end{subequations}}

\title{Inhomogeneous de Sitter Solution with Scalar Field \\
and Perturbations Spectrum}

%\date{\today}
%%%%%%%%%%%%%%%%%%%%%%%%%%%%%%%%
% \email{}
%\affiliation{}
%\superscriptaddress{}
\author{Giovanni Imponente}
%\email{imponente@icra.it}
%\affiliation
\address{Dipartimento di Fisica 
	Universit\`a di Napoli ``Federico II'' and %}
%\affiliation{
INFN-Sezione di Napoli, Italy \\%}
%\affiliation{
ICRA---International Center for Relativistic Astrophysics  \\
Queen Mary, University of London\\
{e-mail: imponente@icra.it}} 

\author{Giovanni Montani}
%\email{montani@icra.it} 
%\affiliation
\address{Dipartimento di Fisica (G9) Universit\`a  
``La Sapienza'' \\
Piazza A.Moro 5,  00185 Roma, Italy \\%}
%\affiliation{
ICRA---International Center for Relativistic Astrophysics\\  
{e-mail: montani@icra.it} }

\maketitle
%\today

\begin{abstract}

We provide an inhomogeneous solution 
concerning the dynamics of a real self 
interacting scalar field minimally coupled 
to gravity in a region of the configuration 
space where it performs a slow rolling 
on a plateau of its potential. 
During the inhomogeneous de Sitter 
phase the scalar field dominant term is 
a function of the
spatial coordinates only. 
This solution specialized nearby the 
FLRW model allows a classical 
origin for the inhomogeneous 
perturbations spectrum.

\end{abstract}

%\keywords{keywords}

%\pacs
\ccode{PACS Nos.: 04.20.Jb, 98.80.Bp-Cq}

\section{General Statements}

A peculiar feature of the inflationary scenario
consists of the violent expansion the Universe
underwent du\-ring the de Sitter 
phase \cite{G81}--\cite{T95};
indeed via such a mechanism the inflationa\-ry 
model provides a satisfactory explanation of 
the so-called \emph{horizons} and \emph{flatness} 
paradoxes by stretching
the inhomogeneities at a very 
large scale \cite{MB95,dB01}.
However, when referred to a (homogeneous 
and isotropic) 
Friedmann--Lemaitre--Robertson--Walker 
(FLRW) model \cite{KT90}, 
the de Sitter phase of the inflatio\-na\-ry 
scenario rules out the small
 inhomogeneous perturbations so strongly, 
 that it makes 
 them unable to become seeds
for the later structures formation \cite{DT94,IM03}.
This picture emerges sharply within 
the inflationary paradigm and it is at 
the ground level of the statements 
according to which 
the cosmological perturbations arise from the 
scalar field quantum fluctuations \cite{STA96}.\\
Though this argument is well settled down 
and results very attractive
even because the predicted quantum spectrum of 
inhomogeneities takes the Harrison-Zeldovich 
form, nevertheless the question 
whether  it is possible, in a more general context,
that classical inhomogeneities survive 
up to a level relevant for the origin of the actual 
Universe large scale structures remains open.\\
Indeed here we investigate the behaviour 
of an inhomogeneous cosmological
model \cite{MAC79,BKL82}
which undergoes a de Sitter phase \cite{STA83,KM02}
and show
how such a general scheme allows the scalar field to 
retain, at the end of the exponential expansion, 
a generic inhomogeneous term to leading order 
(for connected topic see \cite{B87}).\\
Thus our analysis provides relevant information either 
with respect to the morpho\-lo\-gy of an inhomogeneous  
inflationary model, either stating that 
the scalar field
is characterized by an arbitrary spatial function 
which plays the role of its leading order.\\
The model we take into account refers 
to the coupled dynamics of 
an inhomogeneous cosmological model with a 
real self-interacting 
scalar field. The solution concerns 
the phase of evolution when the potential 
associated to the scalar field singles out a plateau 
and the Universe evolution is dominated by 
the effective cosmological constant associated with 
the energy level over the true vacu\-um state of the theory.
We are in condition to neglect the contribution due to the 
ultra-relativistic matter because it would be relevant only 
for higher order terms and becomes more and more
negligible as the exponential expansion develops
(for a discussion of an inflatio\-na\-ry scenario with 
relevant ultra-relativistic matter and different
outcoming behaviour, see \cite{DT94,IM03}).

\section{Inhomogeneous Inflationary Model}

In a synchronous reference, the generic line element
of a cosmological model takes the form
(in units $c=\hbar =1$)
\beq
ds^2 = dt^2 - \gamma_{\alpha\beta}(t, x^\mu)
dx^{\alpha }dx^{\beta } \, , \quad
\alpha,\beta,\mu= 1,2,3 
\label{a}
\eeq
where ${\gamma }_{\alpha \beta }(t, x^{\mu })$ 
is the three-dimensional metric tensor describing 
the geometry of the spatial slices. 
The Einstein equations in the presence 
of a self interacting scalar field 
$\{\phi (t, x^{\mu }), V(\phi)\}$ 
read explicitly  \cite{LK63}
\bseq
\label{einst}
\begin{align} 
& \frac{1}{2} \partial _t k_{\alpha }^{\alpha } + \frac{1}{4}
	k_{\alpha }^{\beta }k_{\beta }^{\alpha } = 
	\chi \left[ - (\partial _t\phi )^2 + V(\phi )\right] 
	\label{b}  \\
& \frac{1}{2}(k^{\beta }_{\alpha ;\beta } - 
	k^{\beta }_{\beta ;\alpha }) = 
	\chi \left(\partial _{\alpha }\phi \,\partial _t \phi \right) 
	\label{c} \\
& \!\!\!\!
	\frac{1}{2\sqrt{\gamma }}\partial _t (\sqrt{\gamma }\,
	k_{\alpha }^{\beta }) + P_{\alpha }^{\beta } = 
	\chi \left[ {\gamma }^{\beta \mu } \,
	\partial _{\alpha }\phi \,\partial _{\mu }\phi + 
 	V(\phi ) \,{\delta }_{\alpha }^{\beta }
	\right] \, ,
	\label{d} 
\end{align}
\eseq
(Einstein constant $\chi = 8\pi G$, 
$G$ being the Newton constant)
where we used the notations 
\beq
\gamma \equiv {\mathrm det} \gamma_{\alpha \beta} \, ,\qquad 
k_{\alpha \beta } \equiv \partial _t{\gamma }_{\alpha \beta } 
\, ,  \qquad k_{\alpha }^{\beta } = 
{\gamma }^{\beta \mu }k_{\alpha \mu } \, .
\label{f}
\eeq
The three-dimensional Ricci tensor
$P_{\alpha }^{\beta } = 
{\gamma }^{\beta \gamma }P_{\alpha \gamma }$ 
is constructed via the metric 
${\gamma }_{\alpha \beta}$ 
which is also used to form the covariant
derivative $(\quad )_{;\alpha }$. 

The dynamics of the scalar field 
$\phi (t, x^{\gamma })$ 
is described by the equation 
\beq 
\partial _{tt}\phi + \frac{1}{2}k_{\alpha }^{\alpha }
\partial _t\phi - {\gamma }^{\alpha \beta }{\phi }_{;\alpha ;\beta } 
+ \frac{dV}{d\phi } = 0 \, ,
\label{m} 
\eeq
coupled  to the Einstein's ones,  
with notation $\partial _{tt} (\quad ) \equiv
\frac{\partial ^2(\quad )}{\partial t^2}$. \\
In what follows we will consider the  
three fundamental statements:
\begin{enumerate}
\renewcommand{\labelenumi}{({\it\roman{enumi}} )}
	\item \label{prima}
	 the three metric tensor is taken in the general
	 factorized form
	 \beq
	 {\gamma }_{\alpha \beta }(t, x^{\mu })=
	 \Gamma^2(t, x^{\mu })\xi_{\alpha \beta }
	 (x^{\mu })
	 \eeq
where $\xi_{\alpha \beta }$ is a generic 
symmetric three-tensor and therefore 
contains six arbitrary functions of the spatial 
coordinates, while $\Gamma$ is to be determined 
by the dynamics.
The inverse metric reads
	 \beq
	 {\gamma }^{\alpha \beta }(t, x^{\mu })=
	 \frac{1}{\Gamma^2(t, x^{\mu })}\xi^{\alpha \beta }
	 (x^{\mu }) \, , 
	 \quad \xi^{\alpha \nu }\xi_{\nu \beta }=
	 \delta^{\alpha}_{ \beta } \, ;
	 \eeq
	\item \label{seconda}
	the self interacting scalar field
	dynamics is described by a potential term 
	which satisfies all the features of an inflationary
	one, i.e. a symmetry breaking confi\-gu\-ration
	characterized by a relevant plateau region; 
	\item \label{terza}
	the inflationary solution is constructed
	under the assumptions
	\bseq
	\label{infla12}
	\bea
  \label{infla1}
  \frac{1}{2}\left(\partial_t \phi\right)^2 &\ll& 
  V\left(\phi\right) \\
  \mid \partial_{tt}\phi \mid &\ll& \mid 
  k^{\alpha}_{\alpha} \, \partial_t\phi \mid
  \label{infla2} \, .
  \eea
  \eseq
\end{enumerate}

Our analysis concerns the evolution of the 
cosmological model when the scalar field slow 
rolls on the plateau and the corresponding 
potential term is described as
	\beq
	\label{pot}
	V(\phi) = \Lambda_0-\lambda U(\phi) \, ,
	\eeq
where $\Lambda_0$ behaves as an effective 
cosmological constant  of the order 
$10^{15}-10^{16}~\mathrm{GeV}$ and $\lambda$
($\ll 1$) is a coupling constant associated
to the perturbation $U(\phi)$.\\
Since the scalar field moves on a 
plateau almost flat, we infer that in 
the lowest order of approximation 
$\phi(t,x^{\gamma})\sim \alpha(x^{\gamma})$ 
(see below (\ref{fi})) and
therefore the potential reduces to 
a space-dependent effective cosmological 
constant 
\beq
\label{pot2}
\Lambda(x^{\gamma})\equiv \Lambda_0
-\lambda U(\alpha(x^{\gamma})) \, .
\eeq
In this scheme the $0-0$ (\ref{b}) and 
$\alpha-\beta$ (\ref{d})
components of the Einstein equations 
reduce, under condition
{\textit{(iii)} and neglecting all the spatial 
gradients, to the simple ones
\bseq
\bea
\label{00}
3\, \partial_{tt}\ln \Gamma+3\, 
(\partial_t \ln \Gamma)^2&=& 
\chi \Lambda(x^{\gamma}) \\
\label{ab}
(\partial_{tt}\ln \Gamma )\delta^{\alpha}_{\beta} +
3\, (\partial_t \ln \Gamma)^2\delta^{\alpha}_{\beta} &=&
\chi \Lambda(x^{\gamma})\delta^{\alpha}_{\beta} \, ,
\eea
\label{00ab}
\eseq
respectively.
A simultaneous solution for $\Gamma$ of both 
equations (\ref{00}) and (\ref{ab}) takes the form
\beq
\label{ve}
\Gamma(x^{\gamma})=\Gamma_0(x^{\gamma})
\exp\left[
\sqrt{
\frac{\chi\Lambda(x^{\gamma})}{3}}(t-t_0)\right] \, ,
\eeq
where $\Gamma_0(x^{\gamma})$ is an integration function
while $t_0$ a given initial instant of time for the 
inflationary scenario.
Under the same assumptions and taking into account 
(\ref{ve}) for $\Gamma$, the scalar field 
equation (\ref{m}) rewrites as 
\beq
\label{vd}
3H(x^{\gamma}) \partial_t \phi - \lambda W(\phi)=0 \, ,
\eeq
where we naturally defined 
\beq
\label{def3}
H(x^{\gamma})=\partial_t \ln \Gamma
=\sqrt{\frac{\chi}{3}\Lambda(x^{\gamma})} \, , 
\qquad W(\phi)=\frac{dU}{d\phi} \, . 
\eeq
We search a solution of the dynamical equation
(\ref{vd}) in the form
\beq
\label{fi}
\phi(t,x^{\gamma})=\alpha(x^{\gamma})+
\beta(x^{\gamma})\left( t-t_0\right) \, .
\eeq
Inserting expression (\ref{fi}) in (\ref{vd})
and considering it to the lowest order, we get the 
relation 
\beq
\label{ba}
3H\beta=\lambda W(\alpha)\, , \qquad
W(\alpha)=\left.\frac{dU}{d\phi}\right|_{\phi=\alpha}.
\eeq
This equation allows to express 
$\beta$ in terms of $\alpha$
\beq
\label{beta}
\beta=\frac{\lambda W(\alpha)}{\sqrt{3\chi 
\Lambda_0 - \lambda U(\alpha)}} \,.
\eeq
Of course the validity of solution (\ref{beta})
takes place in the limit
\beq
\label{lim}
t-t_0 \ll \left| \frac{\alpha}{\beta}
\right| = \left|\frac{\alpha}{ W(\alpha)}
\sqrt{3\chi \frac{\Lambda_0}{\lambda^2} -
\frac{U(\alpha)}{ \lambda}}\,\right|
\eeq
where the ratio $\Lambda_0/\lambda^2$ takes in 
general very large values.\\
The $0-\alpha$ component (\ref{c}) of the 
Einstein equations remains to be solved. 
In view of (\ref{ve})
and (\ref{fi}) through (\ref{beta}) this 
provides the relation 
\beq
\label{rel}
-2 \sqrt{\frac{\chi}{3}}\partial_{\gamma}
\left(\sqrt{\Lambda} \right)=
\chi (\partial_{\gamma}\alpha)\, \beta=
\sqrt{\frac{\chi}{3\Lambda}}\,
\lambda \, \partial_{\gamma}U 
\eeq
or, simplifying easily,
\beq
\label{re}
\partial_{\gamma}
\left( \Lambda + \lambda U\right) =0 \, , 
\eeq
which is reduced to an identity by (\ref{pot2}) 
for $\Lambda(x^{\gamma})$. \\
The validity of the obtained inflationary 
solution is gua\-ran\-teed by considering that all the 
spatial gradients, either of the three-metric
field either of the scalar one, behave 
as $\Gamma^{-2}$ and therefore decay 
exponentially. 

If we take into account the coordinate
characteristic lengths $L$ and $l$
for the inhomogeneity scales 
regarding the functions
$\Gamma_0$ and $\xi_{\alpha \beta}$, 
i.e.
\beq
\partial_{\gamma} \Gamma_0  
\sim \frac{\Gamma_0}{L} \, , \qquad
\partial_{\gamma} \,\xi_{\alpha \beta} 
\sim \frac{\xi_{\alpha \beta}}{l}\, ,
\eeq
respectively, then negligibility of the spatial 
gradients
at the initial instant $t_0$ leads to the 
inequalities for the phy\-sical quantities
\bseq
\bea
\Gamma_0 l &=& l_{\textnormal{phys}} \gg H^{-1} \, , \\
\Gamma_0 L &=& L_{\textnormal{phys}} \gg H^{-1} \, .
\eea
\label{h-1}
\eseq
These conditions state that all the inhomogeneities
have to be much greater then the physical 
horizon $H^{-1}$. 

The assumption made on the  negligibility of the 
spatial gradients at the beginning of the
inflation is required (as well known)
by the existence of the de Sitter phase itself;
however, spatial gradients having a passive 
dynamical role allow to deal with a 
fully  inhomogeneous solution. 
This feature simply means that space point 
dynamically decouple to leading order.

The analysis is completed by stressing that
the condition (\ref{infla1}) becomes 
\beq
\label{inf1}
W^2(\alpha) \ll 
 \chi \left( \frac{\Lambda}{\lambda}\right)^2 \, , 
\eeq 
or equivalently by (\ref{pot2})
\beq
\label{inf11}
\lambda^2\,W^2(\alpha) \ll \chi 
\left( \Lambda_0 - \lambda U(\alpha) \right)^2
\eeq
which, neglecting all terms in $\lambda^2$, 
simply states that the dominant contribution 
in $\Lambda(x^{\gamma})$ is provided by 
$\Lambda_0$, i.e. 
\beq
\label{ua}
\lambda U(\alpha) \ll \Lambda_0 \, ,
\eeq
whereas (\ref{infla2}) is always
naturally satisfied. By other words, 
we get the only important restriction 
on the spatial function $\alpha(x^{\gamma})$
which reads
\beq
\label{ma}
|\alpha| \ll \left|U^{-1}
\left({\Lambda_0}/{\lambda}\right)\right| \, .
\eeq
In order to get a satisfactory 
exponential expansion able to overcome 
the SCM shortcomings, we require 
that in each space point 
the condition
\beq
\label{acca}
H(t_f - t_i) \sim \mathcal{O}(10^2) 
\eeq
holds, where $t_i$ and $t_f$ denote 
the instants when the de Sitter phase starts 
and ends, respectively. 
We may take $t_i\equiv t_0$ and 
$t_f$ must satisfy the inequality 
\beq
\label{ineq}
t_f \ll t^* \equiv t_0 + \left|\frac{\alpha}{\beta} \right|\, .
\eeq
Hence we have 
\beq
\label{ineq1}
H (t_f - t_i) \ll H(t^* -t_0) 
= H \left|\frac{\alpha}{\beta}\right| \, ,
\eeq
or equivalently
\beq
\label{ineq2}
H (t_f - t_i) \ll \frac{\Lambda_0}{\lambda}
\left|\frac{\alpha}{W(\alpha)}\right| \, ,
\eeq
where we made use of (\ref{beta}). 
Being $\Lambda_0/\lambda$ a very large quantity, 
no serious restrictions appear for the e-folding 
of the model.

A fundamental feature of our analysis relies on the 
very general nature of the obtained solution; in fact, 
once sa\-tis\-fied all the dynamical equations, still 
eight arbitrary spatial functions remain, i.e.
six for $\xi_{\alpha \beta}(x^{\gamma})$, and then
$\Gamma_0(x^{\gamma})$, $\alpha(x^{\gamma})$.\\
However, taking into account the possibility 
to choose an arbitrary gauge via the set 
of the spatial coordinates, we have to kill 
three degrees of freedom; hence  
five physically arbitrary functions finally remain:
four corresponding to gravity degrees of freedom
and one related to the scalar field.\\
This picture corresponds exactly to the allowance
of specifying a generic Cauchy problem for the 
gravitational field, on a spatial non-singular
hypersurface, nevertheless one degree of freedom 
of the scalar field is lost against the full 
generality.

\section{Coleman--Weinberg Model}

Let us specify our solution in the case of the
Coleman and Weinberg (CW) zero-temperature potential
\cite{CW73}
\beq
\label{cw}
V(\phi)=\frac{B\sigma^4}{2}
+B\phi^4\left[\ln\left(\frac{\phi^2}{\sigma^2}\right)-\frac12 \right]
	\end{equation}
where $B\simeq 10^{-3}$ is connected to the 
fundamental constants 
of the theory, while
$\sigma\simeq 2\cdot10^{15}\mathrm{GeV}$ 
gives the energy associated with the 
symmetry breaking process.
In the region $\mid \phi\mid\ll \mid \sigma\mid$
the potential (\ref{cw}) approaches a plateau 
behaviour profile similar to (\ref{pot}) 
and acquires the form
\beq
\label{cwa}
V(\phi)\simeq \frac{B\sigma^4}{2} 
-\frac{\lambda}{4}\phi^4 \, , 
\qquad \lambda\simeq80 B\simeq 0.1 \, .
\eeq
This is effectively reducible to (\ref{pot}) by  
\beq
\label{case}
\Lambda_0=\frac{B\sigma^4}{2}\, , \qquad
U(\phi)=\frac{\phi^4}{4} \, , \qquad
 W(\phi)=\phi^3 \, , 
\eeq
and the relations (\ref{beta}) and (\ref{inf1})
rewrite  
\bseq
\begin{align}
\label{betab}
& \beta=\frac{\lambda \alpha^3}{3H} \, , \\
 \label{infl1b}
&\alpha^3\left(\alpha+\sqrt{\frac{8}{3\chi}}\right)
\ll \frac{\Lambda_0}{\lambda}
\simeq \frac{\sigma^4}{160} \, ,
\end{align}
\eseq
respectively. 
The inequality in (\ref{infl1b}) is equivalent
to fulfil the initial assumption
\beq
\label{ini}
\Lambda_0 \gg 
\lambda U(\alpha)\sim\frac{\lambda}{4}\alpha^4 \, ,
\eeq
like as in (\ref{ua}). \\
The restriction (\ref{ma}) reflects 
over the free function $\alpha$ as
\beq
\label{asig}
|\alpha| \ll 
\sqrt[4]{\frac{\Lambda_0}{\lambda}} \sim \sigma \, .
\eeq

\section{Towards FLRW Universe}

The conditions (\ref{h-1}) state that the validity 
of the inhomogeneous inflationary scenario discussed 
in the previous Section requires the 
inhomogeneous scales to be out of the horizon 
when inflation starts. 
The situation is different when treating
the small perturbations to the FLRW case; 
in fact, the negligibility of the spatial curvature 
corresponds to require the radius of curvature 
of the universe to be much greater than the 
physical horizon,  the inhomogeneous terms 
being small in amplitude. 
To this end, let us consider the 
three-metric 
\beq
\gamma_{\alpha \beta} = \Gamma(t,\varphi^{\mu})^2 
\left[h_{\alpha \beta} + (t-t_0)
\delta	\theta_{\alpha \beta}(\varphi^{\mu}) 
	\right] \, , 
	\label{tfw}
\eeq
where $h_{\alpha\beta}$	denotes the FLRW spatial 
part of the three metric 
($\{\varphi^{\mu}\}$
are the three usual angular coordinates) 
and  $\delta	\theta_{\alpha \beta}$ denote 
a small inhomogeneous perturbation. 
The Einstein equations (\ref{einst})
coupled to the scalar field dynamics  (\ref{m})
on the plateau (\ref{cwa}) admit,  
to leading order in the inhomogeneities,
 the solution
\bseq
\begin{align}
&\Gamma=\Gamma_0 e^{H(t-t_0)} \, , \qquad  \qquad
H= H_0 - \frac{\delta	\theta}{6} \, , \\
&%\!\!\!\! 
\phi= \alpha_0 \left[ 1+ 
	\frac{\lambda \alpha_0^2}{3H_0}(t-t_0) \right]
	+ \frac{\delta	\theta}{3 \chi}
  \left[ 1+ 
	\frac{\lambda \alpha_0^2}{H_0}(t-t_0) \right] \\
& \delta	\theta_{\alpha \beta} = 
 	\frac{\delta\theta}{3} h_{\alpha \beta} \, , \qquad \qquad
 H_0 = \sigma^2 \sqrt{\frac{\chi B}{6}}  \, ,
\end{align}
\label{solufrw}
\eseq
where $t_0$ and $ \Gamma_0$ are constants.
The solution (\ref{solufrw}) holds and 
provides the correct e-folding of order 
$\mathcal{O}(10^2)$ when the following 
inequalities take place
\bseq
\begin{align}
& t-t_0 \ll \frac{3 H_0}{\lambda \alpha_0^2} \, , 
\label{di1}\\
& \alpha_0 \ll \mathcal{O} 
	\left( \frac{1}{10}
			\sqrt{\frac{\Lambda_0}{\lambda}}\right) \, ,
			\label{di2}\\
&R_{\textrm{curv}}\equiv \frac{\Gamma_0}{\sqrt{\mathcal{K}}} \gg
	H_0^{-1} \, , \label{di3}\\
& \Gamma_0 l = l_{\textrm{phys}} \gg 	H_0^{-1}\delta \, , \qquad 
	l_{\textrm{phys}} \gg \frac{\delta}{\sqrt{\lambda \alpha_0^3}}\, ;
	\label{di5}
\end{align}
\label{diseq}
\eseq
in (\ref{di3}), $\mathcal{K}$ is the signature of the 
spatial curvature, while $\delta$ 
$(\ll  H_0/100)$
and $l$ in (\ref{di5}) denote the 
cha\-racte\-ristic amplitude and length, 
respectively, of the \textit{arbitrary} function
$\delta \theta$ which is the trace of the 
tensor $\delta \theta_{\alpha \beta}$. 
The inequality (\ref{di1}) ensures that the 
dominant term of the scalar field remains the 
time-independent one during the de Sitter phase; 
inequality (\ref{di2}) allows for an e-folding
of order $\mathcal{O}(10^2)$; 
finally, equations (\ref{di3}) and (\ref{di5})
provide the  negligibility of the spatial gradients 
in the Einstein and scalar field equations, 
respectively. 
When inflation starts, in agreement with 
(\ref{di3}) and (\ref{di5}) the inhomogeneous 
scales can be inside the physical horizon 
$H_0^{-1}$.

The physical implications on the 
density perturbation spectrum of such a nearly 
homogeneous model  rely on the 
dominant behaviour of the potential term 
over the energy 
density $\rho_{\phi}$ associated to the 
scalar field during the de Sitter phase  
and therefore
\beq
\label{drho}
\Delta\equiv 
\left|\frac{\delta\rho_{\phi}}{\rho_{\phi}}\right|	
\sim \left|\frac{d\ln V}{d\phi}\delta\phi \right|
\simeq \left|\frac{\lambda}{\Lambda_0}\,
W(\alpha_0)\,\delta\alpha \right| \, ,
\eeq	 
where $\delta\alpha= \delta\theta/(3\chi)$ 
for our scalar field solution (\ref{solufrw}).
In particular, in the CW
case (\ref{drho}) reduces to 
\beq
\label{cwa1}
\Delta_{CW}\simeq \frac{50}{\sigma^4}
{\alpha_0}^3 \, \frac{\delta\theta}{\chi} \, .
\eeq
However, to get an information
about the problem of computing the physically 
relevant perturbations after the scales re-entry 
in the horizon, we have to deal with 
the gauge invariant quantity $\zeta$ 
\cite{GP82,BST83} which has the form 
\beq
\label{zeta}
\zeta = \frac{\delta \rho}{\rho +p}\cong 
\frac{\delta \theta}{W(\alpha_0)}
\frac{\Lambda_0}{\lambda} 
\eeq
when the perturbations 
leave the horizon and
$\rho + p = (\partial_t \phi)^2$;
in the CW case it reads as 
\beq
\label{zetacw}
\zeta_{CW}=  \frac{\sigma^4}{160}~
\frac{\delta \theta}{{\alpha_0}^3} \, .
\eeq
Since $\zeta$ remains constant during the 
super-horizon evolution of the perturbations, 
then at the re-entry to the causal scale in the 
matter-dominated era, we get
%\beq
$\zeta_{MD} \sim \delta \rho/\rho \sim \zeta_{CW}$.\\
By restoring physical units and assuming 
$\alpha_0 \lesssim  10^{-4}\sigma/\sqrt{hc}$ 
in agreement with (\ref{asig}), then 
 it is required  
$\delta \alpha/\alpha_0 \lesssim 10^{-2}$
in order to obtain perturbations 
$\delta \rho/\rho \sim 10^{-4}$ 
at the  horizon re-entry  
during  the matter-dominated age. \\
Hence the expression (\ref{zetacw}) explains how
the perturbation 
spectrum after the de Sitter phase can 
still arise from classical inhomogeneous 
terms. Indeed, the function 
$\delta\theta(\varphi^{\mu})$ is an arbitrary one
and can be chosen for it a Harrison--Zeldovich
spectrum by assigning its Fourier transform as 
\beq
\label{hz}
\left|\delta \alpha(k)\right|^2\propto 
\frac{\textnormal{const.}}{k^{3}} \, ;
\eeq
such a spectrum has to  hold for 
$k\ll \frac{\Gamma_0}{ H_0^{-1}\delta}$.

Thus, the pre-inflationary 
inhomogeneities of the scalar field
remain almost of the same amplitude 
during the de Sitter phase as a consequence
of the linear form of 
the scalar field solution (\ref{fi}). 
Hence we get that the Harrison--Zeldovich 
spectrum can be a 
pre-inflationary picture of the density perturbations
and it survives to the de Sitter phase, becoming
a classical seed for structure formation.
The existence of such a classical spectrum is not
related with the quantum fluctuations 
of the scalar field whose 
effect is an independent contribution to the 
classical one.

\section{Concluding Remarks}

The merit of our analysis relies on 
having provided a dynamical framework 
within which classical inhomogeneous 
perturbations to a real scalar field 
minimally coupled with gravity can 
survive even after that the de Sitter expansion 
of the universe stretched the geometry;
the key feature underlying this result
consists \textit{(i)} of constructing an 
inhomogeneous model for which the leading
order of the scalar field is provided 
by a spatial function and then  \textit{(ii)}
of showing how the very general case contains 
as a limit a model close to the FLRW one. 

It is relevant to remark that the metric 
tensor (\ref{tfw}) seems of the same form 
as the one considered in \cite{IM03}; 
however in the present paper the function 
$\eta(t)$ appearing in the previous work
is linear in time and does not decay 
exponentially. The different behaviour
relies on the negligibility of the matter 
with respect to the scalar field which 
is at the ground of the present analysis. 
We are here assuming the dynamics of $\eta(t)$
to be driven by the scalar field alone,
instead of by the ultra-relativistic matter.
This situation corresponds to an initial 
conditions for which the scalar field dominates
over the ultra-relativistic matter when 
inflation starts and this is the reason 
for the resulting different issues of the two 
analyses.

\vspace*{6pt}

%%%%%%%%%%%%%%%%%%%%%%%%%%%%%%%%%%%%
%%%%%%%%%%%%%%%%%%%%%%%%%%%%%%%%%%%%%%%%%%%%%%%

\section*{References}

%\vspace{0.5cm} 

%\section*{Acknowledgements}

%\vspace{1cm} 

\end{document}